\documentclass[11pt]{article}
\usepackage{algorithm2e,framed}
\usepackage{amssymb,amsfonts,amsmath}
\usepackage{graphicx}
\newcommand{\Expect}[1]{\mbox{}{\bf{E}}\left[#1\right]}

\newcommand{\FNorm }[1]{\mbox{}\left\|#1\right\|_F  }

\newcommand{\FNormS}[1]{\mbox{}\left\|#1\right\|_F^2}

\newcommand{\TNorm }[1]{\mbox{}\left\|#1\right\|_2  }
\newcommand{\TNormS}[1]{\mbox{}\left\|#1\right\|_2^2}

\newcommand{\XNorm }[1]{\mbox{}\left\|#1\right\|_{\xi}  }

\newcommand{\VTTNorm }[1]{\mbox{}\left|#1\right|_2  }

\newcommand{\setlinespacing}[1]%
           {\setlength{\baselineskip}{#1 \defbaselineskip}}

\newcommand{\abs }[1]{\left|#1\right|}

\newcommand{\BlockDiagk}[1]{\mbox{}\left(%
\begin{array}{cc}
  \Sigma_{k} & \bf{0} \\
  \bf{0} &  \Sigma_{\rho-k}\\
\end{array}\right)}

\newcommand{\BlockDiagkk}[1]{\mbox{}\left(%
\begin{array}{cc}
  \Sigma_{k} & \bf{0} \\
  \bf{0} & \bf{0} \\
\end{array}\right)}

\newcommand{\BlockDiagkrk}[1]{\mbox{}\left(%
\begin{array}{cc}
  \bf{0} & \bf{0} \\
  \bf{0} & \Sigma_{\rho-k} \\
\end{array}\right)}

\newcommand{\BlockDiagkkh}[1]{\mbox{}\left(%
\begin{array}{c}
  \Sigma_{k} \\
  \bf{0} \\
\end{array}\right)}

\newcommand{\BlockDiagkrkh}[1]{\mbox{}\left(%
\begin{array}{c}
  \bf{0} \\
  \Sigma_{\rho-k} \\
\end{array}\right)}

\newtheorem{definition}{Definition}
\newtheorem{lemma}{Lemma}
\newtheorem{theorem}{Theorem}
\newenvironment{Proof}{\noindent {\em Proof:}}

\long\def\killtext#1{}

\begin{document}
\title{
An Improved Approximation Algorithm for the \\
Column Subset Selection Problem
\thanks{A conference proceedings version of this paper appeared in~\cite{BMD09_CSSP_SODA}. This manuscript presents a modified sampling-based algorithm that fixes a bug in Lemma 4.4 of~\cite{BMD09_CSSP_SODA}. This bug also affects the spectral norm bound for the CSSP that was reported in~\cite{BMD09_CSSP_SODA}; the Frobenius norm bound remains unaffected.
}
}

\author{
Christos Boutsidis
\thanks{
Department of Computer Science,
Rensselaer Polytechnic Institute,
Troy, NY,
boutsc@cs.rpi.edu.
}
\and
Michael W. Mahoney
\thanks{
Department of Mathematics,
Stanford University,
Stanford, CA,
mmahoney@cs.stanford.edu.
}
\and
Petros Drineas
\thanks{
Department of Computer Science,
Rensselaer Polytechnic Institute,
Troy, NY, drinep@cs.rpi.edu.
}
}

\date{}
\maketitle

\begin{abstract}
We consider the problem of selecting the ``best'' subset of \emph{exactly $k$ columns} from an $m \times n$ matrix $A$. In particular, we present and analyze a novel two-stage algorithm that
runs in $O(\min\{mn^2,m^2n\})$ time and returns as output an $m \times k$ matrix $C$ consisting of exactly $k$ columns of $A$. In the first stage (the \textit{randomized} stage), the algorithm
randomly selects $\Theta(k \log k)$ columns according to a judiciously-chosen probability distribution that depends on information in the top-$k$ right singular subspace of $A$. In the second stage (the \textit{deterministic} stage), the algorithm applies a deterministic column-selection procedure to select and return exactly $k$ columns from the set of columns selected in the first stage. Let $C$ be the $m \times k$ matrix containing those $k$ columns, let $P_C$ denote the projection matrix onto the span of those columns, and let $A_k$ denote the ``best'' rank-$k$ approximation to the matrix $A$ as computed with the singular value decomposition. Then, we prove that, with probability at least 0.8,
$$
\FNorm{A - P_CA} \leq \Theta\left(k \log^{1/2} k\right) \FNorm{A-A_k}.
$$
This Frobenius norm bound is only a factor of $\sqrt{k \log k}$ worse than the best previously existing
existential result and is roughly $O(\sqrt{k!})$ better than the
best previous algorithmic result (both of Deshpande et
al.~\cite{DRVW06}) for the Frobenius norm version of this Column
Subset Selection Problem. We also prove that, with probability at least 0.8,
$$
\TNorm{A - P_CA}
   \leq \Theta\left(k \log^{1/2} k\right)\TNorm{A-A_k} + \Theta\left(k^{3/4}\log^{1/4}k\right)\FNorm{A-A_k}.
$$
This spectral norm bound is not directly comparable to the best previously existing bounds for the spectral norm version of this Column Subset Selection Problem (such as the ones derived by Gu and Eisenstat in~\cite{GE96}). More specifically, our bound depends on $\FNorm{A-A_k}$, whereas previous results depend on $\sqrt{n-k}\TNorm{A-A_k}$; if these two quantities are comparable, then our bound is asymptotically worse by a $\left(k \log k\right)^{1/4}$ factor.
\end{abstract}
\section{Introduction}
\label{sec:intro}

We consider the problem of selecting the ``best'' set of
\emph{exactly $k$ columns} from an $m \times n$ matrix $A$.
More precisely, we consider the following
\textsc{Column Subset Selection Problem~(CSSP)}:
\begin{definition}
\textbf{(The CSSP)}
Given a matrix $A \in \mathbb{R}^{m \times n}$ and a positive
integer $k$, pick $k$ columns of $A$ forming a matrix $C \in
\mathbb{R}^{m \times k}$ such that the residual $$\XNorm{A - P_C
A}$$ is minimized over all possible ${n \choose k}$ choices for the
matrix $C$. Here, $P_C=CC^+$ denotes the projection onto the
$k$-dimensional space spanned by the columns of $C$ and $\xi = 2\
\mbox{or}\ F$ denotes the spectral norm or Frobenius norm.
\end{definition}
That is, the goal of the CSSP is to find a subset of exactly $k$ columns of $A$
that ``captures'' as much of $A$ as possible, with respect to the
spectral norm and/or Frobenius norm, in a projection sense. The CSSP
has been studied extensively in numerical linear algebra, where it
has found applications in, e.g., scientific computing~\cite{CH92}.
More recently, a relaxation has been studied in theoretical computer
science, where it has been motivated by applications to large
scientific and internet data sets~\cite{DMM08_CURtheory_JRNL}.

\subsection{Complexity of the CSSP}
We briefly comment on the complexity of
the problem. Clearly, in $O(n^k)$ time we can generate all possible matrices $C$ and thus find the optimal solution in $O(n^kmnk)$ time. However, from a practical perspective, in data analysis applications of the CSSP (see Section \ref{sxn:CSSPData}), $n$ is often in the order of hundreds or thousands. Thus, in practice, algorithms whose running time depends exponentially on $k$ are prohibitively slow even if $k$ is, from a theoretical perspective, a constant. Finally, the NP-hardness of the CSSP (assuming $k$ is a function of $n$) is an open problem. Note, though, that a similar problem, asking for the $k$ columns of the $m \times n$ matrix $A$ that maximize the volume of the parallelepiped spanned by the columns of $C$, is provably NP-hard~\cite{AM07_TR}.

\subsection{The CSSP in statistical data analysis}
\label{sxn:CSSPData}

In data applications, where the input matrix $A$ models $m$ objects
represented with respect to $n$ features, the CSSP corresponds to
unsupervised feature selection. Standard motivations for feature
selection include facilitating data visualization, reducing training
times, avoiding overfitting, and facilitating data understanding.

Consider, in particular, Principal Components Analysis (PCA), which
is the predominant linear dimensionality reduction technique, and which
has been widely applied on datasets in all scientific domains, from
the social sciences and economics, to biology and chemistry. In
words, PCA seeks to map or embed data points from a high dimensional
Euclidean space to a low dimensional Euclidean space while keeping all the relevant
linear structure intact. PCA is an unsupervised dimensionality
reduction technique, with the sole input parameters being the
coordinates of the data points and the number of dimensions that
will be retained in the embedding (say $k$), which is typically a
constant independent of $m$ and $n$; often it is $k \ll \{m,n\}$ too. Data analysts often seek a
subset of $k$ actual features (that is, $k$ actual columns, as opposed to
the $k$ eigenvectors or eigenfeatures
returned by PCA) that can accurately reproduce the structure derived
by PCA. The CSSP is the obvious optimization problem associated with
such unsupervised feature selection tasks.

We should note that similar formulations
appeared in~\cite{Krzanowski87,SDDO03,WS05,ZL07,Mao05,BHG03}.
In addition, applications of such ideas include: ($i$)~\cite{SXZF07_cmdmatrix}, where
a ``compact CUR matrix decomposition'' was applied to static and
dynamic data analysis in large sparse graphs; ($ii$)~\cite{MMD06,MMD08_tensorCUR_SIMAX,DKR02},
where these ideas were used for compression and classification of hyperspectral medical data
and the reconstruction of missing entries from recommendation systems data in order to make
high-quality recommendations; and ($iii$)~\cite{Paschou07b}, where the
concept of ``PCA-correlated SNPs'' (Single Nucleotide
Polymorphisms) was introduced and applied to classify individuals from throughout
the world without the need for any prior ancestry
information.
See~\cite{BMD08_CSSP_KDD} for a detailed evaluation of
our main algorithm as an unsupervised feature selection
strategy in three application domains of modern statistical data
analysis (finance, document-term data, and genetics).

\subsection{Our main results}

We present a novel two-stage algorithm for the CSSP.
This algorithm is presented in
detail in Section~\ref{sxn:main_alg} as Algorithm~\ref{alg:main}.
In the first stage of this algorithm (the
\textit{randomized stage}), we randomly select $\Theta(k \log k)$ columns
of $V_k^T$, i.e., of the transpose of the $n \times k$ matrix
consisting of the top $k$ right singular vectors of $A$, according
to a judiciously-chosen probability distribution that depends on
information in the top-$k$ right singular subspace of $A$. Then, in
the second stage (the \textit{deterministic stage}), we apply a
deterministic column-selection procedure to select exactly $k$
columns from the set of columns of $V_k^T$ selected by the first
stage. The algorithm then returns the corresponding $k$ columns of
$A$. In Section~\ref{sxn:main_proof} we prove the following theorem.

\begin{theorem}
\label{thm:main} There exists an algorithm (the two-stage
Algorithm~\ref{alg:main}) that approximates the solution to the CSSP.
This algorithm takes as input an $m \times n$ matrix $A$ and a positive integer $k$; it runs in
$O(\min\{mn^2,m^2n\})$ time; and it returns as output an $m \times k$
matrix $C$ consisting of exactly $k$ columns of $A$ such that with
probability at least $0.8$:
\begin{eqnarray*}
\TNorm{A - P_CA}
   &\leq& \Theta\left(k \log^{1/2} k\right)\TNorm{A-A_k} + \Theta\left(k^{3/4}\log^{1/4}k\right)\FNorm{A-A_k},  \\
\FNorm{A - P_CA}
   &\leq&
          \Theta\left(k \log^{1/2} k\right) \FNorm{A-A_k}.
\end{eqnarray*}
Here, $P_C=CC^+$ denotes a projection onto the column span of the matrix
$C$, and $A_k$ denotes the best rank-$k$ approximation to the
matrix $A$ as computed with the singular value decomposition.
\end{theorem}
Note that we can trivially boost the success probability in the above theorem
to $1-\delta$ by repeating the algorithm
$O\left(\log\left(1/\delta\right)\right)$ times. Note also that the
running time of our algorithm is linear in the larger of the
dimensions $m$ and $n$, quadratic in the smaller one, and
independent of $k$. Thus, it is practically useful and efficient.

To put our results into perspective, we compare them to the best existing results for the CSSP. Prior work provided bounds of the form
\begin{eqnarray}
\label{eqn:priorbound}\XNorm{A - P_CA}
   &\leq& p(k,n) \XNorm{A-A_k},
\end{eqnarray}
where $p(k,n)$ is a polynomial on $n$ and $k$. For $\xi = 2$, i.e., for the spectral norm, the best previously-known bound for approximating the CSSP is $p(k,n) =\Theta\left(\sqrt{k(n-k)+1}\right)$~\cite{GE96}, while for $\xi = F$, i.e., for the Frobenius norm, the best bound is $p(k,n) = \sqrt{(k+1)!}$~\cite{DRVW06}. Both results are algorithmically efficient, running in time polynomial in all three parameters $m$, $n$, and $k$. The former runs in $O(mnk \log n)$ time and the latter runs in $O(mnk + kn)$ time. Our approach provides an algorithmic bound for the Frobenius norm version of the CSSP that is roughly $O(\sqrt{k!})$ better than the best previously-known algorithmic result. It should  be noted that \cite{DRVW06} also proves that by exhaustively testing all ${n \choose k}$ possibilities for the matrix $C$, the best one will satisfy eqn. (\ref{eqn:priorbound}) with $p(k,n) = \sqrt{k+1}$. Our algorithmic result is only $O(\sqrt{k\log k})$ worse than this existential result. A similar existential result for the spectral norm version of the CSSP is proved in \cite{HP92} with $p(k,n) = \sqrt{k(n-k)+1}$. Our spectral norm bound depends on  $\Theta\left(k^{3/4}\log^{1/4}k\right)\FNorm{A-A_k}$. In a worst case setting (e.g., when all the bottom $n-k+1$ singular values of $A$ are equal) this quantity is upper bounded by
$\Theta\left(\left(n-k\right)^{1/2}k^{3/4}\log^{1/4}k\right)\TNorm{A-A_k}$. This is worse than the best results for the spectral norm version of the CSSP by a factor of $\Theta\left(k^{1/4}\log^{1/4} k\right)$.

Finally, we should emphasize that a novel feature of the algorithm that we present in this paper is that it
combines in a nontrivial manner recent algorithmic developments in
the theoretical computer science community with more traditional
techniques from the numerical linear algebra community in order to
obtain improved bounds for the CSSP.
\section{Background and prior work}
\label{sxn:prior_work}

\subsection{Notation and linear algebra}

First, recall that the $\Theta$-notation can be used to denote an asymptotically tight bound: $f(n) \in \Theta(g(n))$ or $f(n) = \Theta(g(n))$ if there exist positive constants $c_1$, $c_2$, and
$n_0$ such that $0 \le c_1 g(n) \le f(n) \le c_2 g(n)$ for all $n \ge n_0$. This is similar to the way in which the big-$O$-notation can be used to denote an asymptotic upper bound: $f(n) = O(g(n))$ if there exist positive constants $c$ and $n_0$ such that $0 \le f(n) \le c g(n)$ for all $n \ge n_0$.

Let $[n]$ denote the set $\{1,2,\ldots,n\}$. For any matrix $A \in \mathbb{R}^{m \times n}$, let $A_{(i)}, i \in [m]$ denote the $i$-th row of $A$ as a row vector, and let $A^{(j)}, j \in [n]$ denote the $j$-th column of $A$ as a column vector. In addition, let $ \FNormS{A} = \sum_{i,j} A_{ij}^2 $ denote the square of its Frobenius norm, and let $\TNorm{A} = \sup_{x\in \mathbb{R}^n,\ x\neq 0}
\VTTNorm{Ax}/\VTTNorm{x}$ denote its spectral norm. If $A \in \mathbb{R}^{m \times n}$, then the Singular Value Decomposition (SVD) of $A$ can be written as
\begin{eqnarray}
\label{svdA}
\nonumber
A  &=& U_A \Sigma_A V_A^T         \\
\nonumber
         &=& \left(\begin{array}{cc}
             U_{k} & U_{\rho-k}
          \end{array}
    \right)
    \left(\begin{array}{cc}
             \Sigma_{k} & \bf{0}\\
             \bf{0} & \Sigma_{\rho - k}
          \end{array}
    \right)
    \left(\begin{array}{c}
             V_{k}^T\\
             V_{\rho-k}^T
          \end{array}
    \right).
\end{eqnarray}
In this expression, $\rho \leq \min\{m,n\}$ denotes the rank of $A$,
$U_A \in \mathbb{R}^{m \times \rho}$ is an orthonormal matrix,
$\Sigma_A$ is a $\rho \times \rho$ diagonal matrix, and $V_A \in
\mathbb{R}^{n \times \rho}$ is an orthonormal matrix. Also,
$\Sigma_k$ denotes the $k \times k$ diagonal matrix containing the
top $k$ singular values of $A$, $\Sigma_{\rho-k}$ denotes the
$\left(\rho-k\right) \times \left(\rho - k\right)$ matrix containing
the bottom $\rho-k$ singular values of $A$, $V_{k}$ denotes the $n
\times k$ matrix whose columns are the top $k$ right singular
vectors of $A$, and $V_{\rho - k}$ denotes the $n \times \left(\rho
- k\right)$ matrix whose columns are the bottom $\rho - k$ right
singular vectors of $A$, etc.

The $m \times k$ orthogonal matrix $U_k$ consisting of the top $k$
left singular vectors of $A$ is the ``best'' set of $k$ linear
combinations of the columns of $A$, in the sense that $ A_k = P_{U_k} A
=U_k\Sigma_kV_k^T $ is the ``best'' rank $k$ approximation to $A$.
Here, $P_{U_k} =  U_k U_k^T$ is a projection onto the $k$-dimensional
space spanned by the columns of $U_k$. In particular, $A_k$
minimizes $\XNorm{A-A^{\prime}}$, for both $\xi = 2\ \mbox{and}\ F$,
over all $m \times n$ matrices $A^{\prime}$ whose rank is at most
$k$.
We also denote $A_{\rho - k} =
U_{\rho - k} \Sigma_{\rho - k} V_{\rho - k}^T$.
We will use the notation $\XNorm{\cdot}$ when writing an
expression that holds for both the spectral and the Frobenius norm.
We will subscript the norm by $2$ and $F$ when writing expressions
that hold for one norm or the other. Finally, the Moore-Penrose
generalized inverse, or pseudoinverse, of $A$, denoted by $A^+$, may
be expressed in terms of the SVD as $A^+=V_A \Sigma_{A}^{-1}U_A^T$.

Finally, we will make frequent use of the following fundamental result from probability theory, known as Markov's inequality~\cite{MotwaniRaghavan95}. Let $X$ be a random variable assuming non-negative values with expectation $\Expect{X}$. Then, for all $t > 0$,
$X \leq t \cdot \Expect{X}$
with probability at least $1-t^{-1}$. We will also need the so-called union bound. Given a set of probabilistic events ${\cal E}_1,{\cal E}_2,\ldots,{\cal E}_{n}$ holding with respective probabilities $p_1,p_2,\ldots,p_n$, the probability that all events hold (a.k.a., the probability of the union of those events) is upper bounded by $\sum_{i=1}^n p_i$.

\subsection{Related prior work}

Since solving the CSSP exactly
is a hard combinatorial optimization problem, research
has historically focused on computing approximate solutions to it.
Since $\XNorm{A -
A_k}$ provides an immediate lower bound for $\XNorm{A - P_CA}$,
for $\xi=2,F$ and for any choice of $C$, a large number of
approximation algorithms have been proposed to select a subset of
$k$ columns of $A$ such that the resulting matrix $C$ satisfies
\begin{equation*}
\XNorm{A - A_k}
   \leq \XNorm{A - P_CA}
   \leq p(k,n) \XNorm{A - A_k}
\end{equation*}
for some function $p(k,n)$. Within the numerical linear algebra
community, most of the work on the CSSP has focused on spectral norm
bounds and is related to the so-called
\textsc{Rank Revealing QR (RRQR) factorization}:
\begin{definition}
\label{def:rrqr}
\textbf{(The RRQR factorization)}
Given a matrix $A \in R ^ {m \times n}$  ($m \geq n$) and an integer $k$
($k \leq n$), assume partial $QR$ factorizations of the form:
\begin{eqnarray*}
\label{eq:rrqr}
A \Pi = Q R
      = Q\left(
            \begin{array}{cc}
                 R_{11} & R_{12} \\
                 0  & R_{22}
            \end{array}
         \right)   ,
\end{eqnarray*}
where $Q \in R^{m \times n}$ is an orthonormal matrix, $R \in R^{n
\times n}$ is upper triangular, $R_{11} \in R^{k \times k}$, $R_{12}
\in R^{k \times (n-k)}$, $R_{22} \in R^{(n-k) \times (n-k)}$, and
$\Pi \in R^{n \times n}$ is a permutation matrix. The above
factorization is called a RRQR factorization if it satisfies
\begin{eqnarray*}
\label{eq:rrqrb}
\frac{\sigma_{k}(A)} { p_1(k,n)} \leq &\sigma_{min}(R_{11})& \leq
\sigma_{k}(A)\\
\sigma_{k+1}(A) \leq &\sigma_{max}(R_{22})& \leq p_2(k,n)
\sigma_{k+1}(A),
\end{eqnarray*}
where $p_1(k,n)$ and $p_2(k,n)$ are functions bounded by low degree
polynomials in $k$ and $n$.
\end{definition}

The work of Golub on pivoted $QR$
factorizations~\cite{Golub65} was followed by much research addressing
the problem of constructing an efficient RRQR factorization. Most
researchers improved RRQR factorizations by focusing on improving
the functions $p_1(k,n)$ and $p_2(k,n)$ in
Definition~\ref{def:rrqr}. Let $\Pi_k$ denote the first $k$ columns
of a permutation matrix $\Pi$. Then, if $C=A\Pi_k$ is an $m \times
k$ matrix consisting of $k$ columns of $A$, it is straightforward to
prove that
\begin{equation*}
\XNorm{ A - P_CA }  = \XNorm{ R_{22}},
\end{equation*}
for both $\xi=2,F$.
Thus, in particular, when applied to the spectral norm, it follows that
\begin{equation*}
\TNorm{ A - P_CA }
 \leq  p_2(k,n) \sigma_{k+1}(A)
 = p_2(k,n)\TNorm{A-A_k}  ,
\end{equation*}
i.e., any algorithm that constructs an RRQR factorization of the
matrix $A$ with provable guarantees also provides provable
guarantees for the CSSP. See Table~\ref{tb:priorwork} for a summary
of existing results, and see~\cite{FL06_DRAFT} for a survey and an empirical
evaluation of some of these algorithms. More recently,
\cite{MRT06_TR2,WLRT07_TR} proposed random-projection type algorithms that
achieve the same spectral norm bounds as prior work while improving
the running time.

\begin{table*}
\begin{center}
\begin{tabular}{|l|l l|l|l|}
\hline \hline
  \textbf{Method} & \textbf{Reference }  &                       &
\textbf{p(k,n)}     & \textbf{Time}                             \\
  \hline \hline
  Pivoted QR   &  \text{[Golub, 1965]} & \cite{Golub65} &  $\sqrt{(n-k)}
2^{k}$           &        $O(mnk)$                      \\
  \hline
  High RRQR  & \text{[Foster, 1986]} & \cite{Fos86}                 &
$\sqrt{n(n-k)}  2^{n-k}    $ &$O(mn^2)$    \\
  \hline
  High RRQR  & \text{[Chan, 1987]} & \cite{Cha87}                  &
$\sqrt{n(n-k)}  2^{n-k}    $ & $O(mn^2)$ \\
  \hline
  RRQR   & \text{[Hong and Pan, 1992]} &
\cite{HP92}                      & $\sqrt{k(n-k)+ k} $ & -    \\
  \hline
  Low RRQR   & \text{[Chan and Hansen, 1994]} &
\cite{CH94}                   & $\sqrt{(k+1)n}  2^{k+1}  $ & $O(mn^2)$\\
  \hline
  Hybrid-I RRQR       & \text{[Chandr. and Ipsen, 1994]} &
\cite{CI94}                  & $\sqrt{(k+1)(n-k)}         $ & - \\
  Hybrid-II RRQR      &      &
\cite{CI94}                                                   &
$\sqrt{(k+1)(n-k)}          $ & - \\
  Hybrid-III RRQR     &      &
\cite{CI94}                                                   &
$\sqrt{(k+1)(n-k)}          $ & - \\

  \hline
  Algorithm 3   & \text{[Gu and Eisenstat, 1996]} &
\cite{GE96}               & $\sqrt{k(n-k)+1}           $  & -  \\
  Algorithm 4   &          &
\cite{GE96}                                      &
$\sqrt{f^2k(n-k)+1}         $ & $O(kmn\log_{f}(n))$ \\
  \hline
  DGEQPY        & \text{[Bischof and Orti, 1998]} &
\cite{BQ98a}               & $O(\sqrt{(k+1)^2(n-k)})   $ & - \\
  DGEQPX        &                 & \cite{BQ98a}
& $O(\sqrt{(k+1)(n-k)})    $   & -   \\
  \hline
    SPQR   & \text{[Stewart, 1999]}       &     \cite{Ste99}       & - &
-   \\
  \hline
  PT Algorithm 1       & \text{[Pan and Tang, 1999]} & \cite{PT99}
& $O(\sqrt{(k+1)(n-k)})$    & - \\
  PT Algorithm 2       &                   & \cite{PT99}
& $O(\sqrt{(k+1)^2(n-k)})$ & - \\
  PT Algorithm 3       &                   & \cite{PT99}
& $O(\sqrt{(k+1)^2(n-k)}) $ & - \\
  \hline
  Pan Algorithm 2       & \text{[Pan, 2000]} &
\cite{Pan00}                           & $O(\sqrt{k(n-k)+ 1})$   &
-  \\
  \hline
\end{tabular}
\caption{Deterministic RRQR algorithms for the CSSP. A dash implies that either the authors do not provide a running time bound or the algorithm depends exponentially on $k$.
(In addition, $m \geq n$ and $f \geq 1$ for this table.)
\label{tb:priorwork}
}
\end{center}
\end{table*}


Within the theoretical computer science community, much work has
followed that of Frieze, Kannan, and Vempala~\cite{FKV98} on
selecting a small subset of representative columns of $A$, forming a
matrix $C$, such that the projection of $A$ on the subspace spanned
by the columns of $C$ is as close to $A$ as possible. The algorithms
from this community are randomized, which means that they come with
a failure probability, and focus mainly on the Frobenius norm. It is
worth noting that they provide a strong tradeoff between the number
of selected columns and the desired approximation accuracy. A
typical scenario for these algorithms is that the desired
approximation error (see $\epsilon$ below) is given as input, and
then the algorithm selects the minimum number of appropriate columns
in order to achieve this error. One of the most relevant results for
this paper is a bound of~\cite{DRVW06}, which states that there
exist exactly $k$ columns in any $m \times n$ matrix $A$ such that
\begin{equation*}
\label{eqn:random} \FNorm{A - CC^+A} \leq \sqrt{k+1}\FNorm{A-A_k}.
\end{equation*}
Here, $C$ contains exactly $k$ columns of $A$. The only known
algorithm to find these $k$ columns is to try all ${n \choose k}$
choices and keep the best.
This existential result relies on the so-called
volume sampling method~\cite{DRVW06,DV06}. In~\cite{DV06}, an
adaptive sampling method is used to approximate the volume sampling
method and leads to an $O(mnk+kn)$ algorithm which finds $k$ columns
of $A$ such that
\begin{equation*}
\label{eqn:random1} \FNorm{A - CC^+A} \leq
\sqrt{(k+1)!}\FNorm{A-A_k}.
\end{equation*}
As mentioned above, much work has also considered algorithms choosing
slightly more than $k$ columns. This relaxation provides significant
flexibility and improved error bounds. For example, in~\cite{DV06},
an adaptive sampling method leads to an $O\left(mn\left(k/\epsilon^2
+ k^2 \log k\right)\right)$ algorithm, such that
\begin{equation*}
\label{eqn:random2} \FNorm{A - CC^+A} \leq
\left(1+\epsilon\right)\FNorm{A-A_k}
\end{equation*}
holds with high probability for some matrix $C$ consisting of
$\Theta\left(k/\epsilon^2 + k^2\log k\right)$ columns of $A$.
Similarly, in~\cite{DMM06b,DMM08_CURtheory_JRNL}, Drineas, Mahoney, and Muthukrishnan
leverage the subspace sampling method to give an $O(\min\{mn^2,m^2n\})$
algorithm such that
\begin{equation}
\label{eqn:random3} \FNorm{A - CC^+A} \leq
(1+\epsilon)\FNorm{A-A_k}
\end{equation}
holds with high probability if $C$ contains $\Theta(k \log
k/\epsilon^2)$ columns of $A$.
\section{A two-stage algorithm for the CSSP}
\label{sxn:main_alg}

In this section, we present and describe Algorithm~\ref{alg:main}, our main algorithm for approximating the solution to the CSSP. This algorithm takes as input an $m \times n$ matrix $A$ and a rank parameter $k$. After an initial setup, the algorithm has two stages: a randomized stage and a deterministic stage. In the \textit{randomized stage}, a randomized procedure is run to select $\Theta\left(k \log k\right)$ columns from the $k \times n$ matrix $V_k^T$, i.e., the transpose of the matrix containing the top-$k$ right singular vectors of $A$. The columns are chosen by randomly sampling according to a judiciously-chosen nonuniform probability distribution that depends on information in the top-$k$ right singular subspace of $A$. Then, in the \textit{deterministic stage}, a deterministic procedure is employed to select exactly $k$ columns from the $\Theta\left(k \log k\right)$ columns chosen in the randomized stage. The algorithm then outputs exactly $k$ columns of $A$ that correspond to those columns chosen from $V_k^T$. Theorem~\ref{thm:main} states that the projection of $A$ on the subspace spanned by these $k$ columns of $A$ is (up to bounded error) close to the best rank $k$ approximation to~$A$.

\subsection{Detailed description of our main algorithm}

In more detail, Algorithm~\ref{alg:main} first computes a probability distribution $p_1,p_2,\ldots,p_n$ over the set $\{1,\ldots,n\}$, i.e., over the columns of $V_k^T$, or equivalently over the columns of $A$. The probability distribution depends on information in the top-$k$ right singular subspace of $A$. In particular, for all $i \in [n]$, define
\begin{align}
\label{eqn:sampling_probs} p_i = \frac{\frac 1 2
\TNormS{\left(V_{k}\right)_{(i)}}}{\sum_{j=1}^n
\TNormS{\left(V_{k}\right)_{(j)}}}+\frac{\frac 1 2
\TNormS{\left(\Sigma_{\rho-k}V_{\rho-k}^T\right)^{(i)}}}{\sum_{j=1}^n
\TNormS{\left(\Sigma_{\rho-k}V_{\rho-k}^T\right)^{(j)}}}   ,
\end{align}
and note that $p_i\ge0$, for all $i\in [n]$, and that $\sum_{i=1}^n p_i = 1$. We will describe the computation of probabilities of this form below.

In the \textit{randomized stage}, Algorithm~\ref{alg:main} employs the following randomized column selection algorithm to choose $\Theta(k \log k)$ columns from $V_k^T$ to pass to the second stage. Let $c$ assume the value of eqn.~(\ref{eqn:valueofc}). In $c$ independent identically distributed (i.i.d.) trials, the algorithm chooses a column of $V_k^T$ where in each trial the $i$-th column of $V_k^T$ is kept with probability $p_i$. Additionally, if the $i$-th column is kept, then a scaling factor equal to $1/\sqrt{cp_i}$ is kept as well. Thus, at the end of this process, we will be left with $c$ columns of $V_k^T$ and their corresponding scaling factors. Notice that due to random sampling in i.i.d. trials with replacement we might keep a particular column more than once.

In order to conveniently represent the $c$ selected columns and the associated scaling factors,
we will use the following sampling matrix formalism. First, define
an $n \times c$ sampling matrix $S_1$ as follows: $S_1$ is
initially empty; at each of the $c$ i.i.d. trials, if the $i$-th column of
$V_k^T$ is selected by the random sampling process, then $e_i$ (an
$n$-vector of all-zeros, except for its $i$-th entry which is set
to one) is appended to $S_1$. Next, define the $c \times c$ diagonal rescaling matrix $D_1$ as follows: if the
$i$-th column of $V_k^T$ is selected, then a diagonal entry of
$D_1$ is set to $1/\sqrt{cp_i}$. Thus, we may view the randomized stage as outputting the matrix $V_k^TS_1D_1$ consisting of a small number of rescaled columns of $V_k^T$, or
simply as outputting $S_1$ and $D_1$.

In the \textit{deterministic stage}, Algorithm~\ref{alg:main}
applies a deterministic column selection algorithm to the output of
the first stage in order to choose \emph{exactly $k$ columns} from
the input matrix $A$. To do so, we run the Algorithm 4 of~\cite{GE96} (with the parameter $f$ set to $\sqrt{2}$) on the $k \times c$ matrix $V_k^T S_1 D_1$, i.e., the column-scaled version of the columns of $V_k^T$ chosen in the first stage. Thus, a matrix $V_k^T S_1D_1S_2$ is formed, or equivalently, in the sampling matrix formalism described previously, a new matrix
$S_2$ is constructed. Its dimensions are $c \times k$, since
it selects exactly $k$ columns out of the $c$ columns
returned after the end of the randomized stage. The algorithm then
returns the corresponding $k$ columns of the original matrix $A$,
i.e., after the second stage of the algorithm is complete, the $m
\times k$ matrix $C = AS_1 S_2$ is returned as the final output.

\begin{algorithm}[ht]

\begin{framed}

\noindent \textbf{Input:} $m \times n$ matrix $A$, integer $k$.

\noindent \textbf{Output:} $m \times k$ matrix $C$ with $k$
columns of $A$.
\\
\begin{enumerate}
\item
\textbf{Initial setup:}
\begin{itemize}
\item
Compute the top $k$ right singular vectors of $A$, denoted by
$V_k$.
\item
Compute the sampling probabilities $p_i$, for $i \in [n]$, using
eqn.~(\ref{eqn:sampling_probs}) or eqn.~(\ref{eq:probexpression}).
\item
Let
\begin{equation}\label{eqn:valueofc}
c = 1600c_0^2 k \log \left(800c_0^2 k\right) = \Theta(k \log k).
\end{equation}
(Here $c_0$ is the unspecified constant of Theorem~\ref{thm:theorem7correct}.)
\end{itemize}
\item
\textbf{Randomized Stage:}
\begin{itemize}
\item
For $t=1,\ldots,c$ (i.i.d. trials) select an integer from $\left\{1,2,\ldots,n\right\}$ where the probability of selecting $i$ is equal to $p_i$. If $i$ is selected, keep the scaling
factor $1/\sqrt{cp_i}$.
\item
Form the sampling matrix $S_1$ and the rescaling matrix $D_1$ (see
text).
\end{itemize}
\item
\textbf{Deterministic Stage:}
\begin{itemize}
\item
Run Algorithm 4, page 853 of~\cite{GE96} (with the parameter $f$ set to $\sqrt{2}$) on the matrix $V_k^TS_1D_1$ in order to select
exactly $k$ columns of $V_k^TS_1D_1$, thereby forming the sampling matrix $S_2$ (see text).
\item
Return the corresponding $k$ columns of $A$, i.e., return $C =
AS_1S_2$.
\end{itemize}

\end{enumerate}
\caption{A two-stage algorithm for the CSSP.} \label{alg:main}
\end{framed}
\end{algorithm}

\subsection{Running time analysis}
We now discuss the running time of our algorithm. Note that manipulating the probability distribution of eqn.~(\ref{eqn:sampling_probs})
yields:
\begin{equation}
\label{eq:probexpression}
p_i =
\frac{\TNormS{\left(V_{k}\right)_{(i)}}}{2k}
    + \frac{\TNormS{\left(A\right)^{(i)}} -
            \TNormS{\left(AV_kV_k^T\right)^{(i)}}}
           {2\left(\FNormS{A}-\FNormS{AV_kV_k^T}\right)}   .
\end{equation}
Thus, knowledge of $V_k$, i.e., the $n \times k$ matrix consisting of the top-$k$ right singular vectors of $A$, suffices to compute the $p_i$'s. By eqn.~(\ref{eq:probexpression}), $O(\min\{mn^2,m^2n\})$ time suffices for our theoretical analysis. In practice iterative algorithms could be used to speed up the algorithm. Note also that in order to obtain the Frobenius norm bound of Theorem~\ref{thm:main}, our theoretical analysis holds if the sampling probabilities are of the form:
\begin{eqnarray}
\label{eqn:sampling_probs_firstterm}
p_i = \TNormS{\left(V_{k}\right)_{(i)}}/k.
\end{eqnarray}
That is, the Frobenius norm bound of Theorem~\ref{thm:main} holds even if the second term in the sampling probabilities of eqns.~(\ref{eqn:sampling_probs}) and (\ref{eq:probexpression})
is omitted.

Finally, we briefly comment on a technical constraint of Algorithm 4 of~\cite{GE96}. This algorithm assumes that its input matrix has at least as many rows as columns. However, in our approach, we will apply it on the $k \times c$ matrix $V_k^TS_1D_1$, which clearly has fewer rows than columns. Thus, prior to applying the aforementioned algorithm, we first pad $V_k^T S_1 D_1$ with $c-k$ all-zero rows, thus making it a square matrix. Let $\Omega = V_k^T S_1 D_1$ and let $\tilde{\Omega}$ be the $c \times c$ matrix after the padding. Eqn.~(8) in Theorem 3.2 of~\cite{GE96} (with $i$ set to $k$ and $f$ set to $\sqrt{2}$) implies that $\sigma_k(\tilde{\Omega} S_2) \geq \sigma_k(\tilde{\Omega}) / \left(\sqrt{1 + 2k(c - k)}\right)$. Clearly, $\sigma_k(\tilde{\Omega} S_2) = \sigma_k(\Omega S_2)$ and $\sigma_k(\tilde{\Omega}) = \sigma_k(\Omega)$. Overall, we get, 
$$ \sigma_k(V_k^T S_1 D_1 S_2) \geq \frac{\sigma_k\left(V_k^T S_1 D_1\right)} {\sqrt{1 + 2k(c - k)}}, $$
which is the only guarantee that we need in the deterministic step (see Lemma~\ref{lemma:sigmabound}). The running time of the deterministic stage of Algorithm~\ref{alg:main} is $O(c^2 k \log \sqrt{c})$ time, since the (padded) matrix $V_k^TS_1D_1$ has $c$ columns and rows.

An important open problem would be to identify other suitable importance sampling probability distributions that avoid the computation of a basis for the top-$k$ right singular subspace.

\subsection{Intuition underlying our main algorithm}

Intuitively, we achieve improved bounds for the CSSP because we apply the deterministic algorithm to a lower dimensional matrix (the matrix $V_k^TS_1D_1$ with $\Theta\left(k \log k\right)$ columns, as opposed to the matrix $A$ with $n$ columns) in which the columns are ``spread out'' in a ``nice'' manner. To see this, note that the probability distribution of eqn.~(\ref{eqn:sampling_probs_firstterm}), and thus one of the two terms in the probability distribution of eqns.~(\ref{eqn:sampling_probs}) or (\ref{eq:probexpression}),
equals (up to scaling) the diagonal elements of the projection matrix onto the span of the top-$k$ right singular subspace. In diagnostic regression analysis, these quantities have a natural
interpretation in terms of \emph{statistical leverage}, and thus they have been used extensively to identify ``outlying'' data points~\cite{ChatterjeeHadi88}. Thus, the importance sampling probabilities that we employ in the randomized stage of our main algorithm provide a bias toward more ``outlying'' columns, which then provide a ``nice'' starting point for the deterministic stage of our
main algorithm. This also provides intuition as to why using importance sampling probabilities of the form of eqn.~(\ref{eqn:sampling_probs_firstterm}) leads to relative-error low-rank matrix approximations~\cite{DMM06b,DMM08_CURtheory_JRNL}.

\section{Proof of Theorem \ref{thm:main}}
\label{sxn:main_proof}

In this section, we provide a proof of Theorem~\ref{thm:main}. We start with an outline of our proof, pointing out conceptual improvements that were necessary in order to obtain improved bounds. An important condition in the first phase of the algorithm is that when we sample columns from the $k \times n$ matrix $V_k^T$, we obtain a $k \times c$ matrix $V_k^TS_1D_1$ that does not lose any rank. To do so, we will apply a result from matrix perturbation theory to prove that if $c = \Theta(k \log k)$ (see eqn.~(\ref{eqn:valueofc})) then $\abs{\sigma_{k}^2\left(V_{k}^T S_1D_1\right)-1} \leq 1/2$. (See Lemma~\ref{lemma:rank} below.) Then, under the assumption that $V_k^TS_1D_1$ has full rank, we will prove that the $m \times k$ matrix $C$ returned by the algorithm will satisfy:
\begin{equation*}
\XNorm{A - P_CA}
   \leq \XNorm{A - A_k} \\
      + \sigma_k^{-1}\left(V_{k}^T S_1D_1S_2\right)
        \XNorm{\Sigma_{\rho - k} V_{\rho - k}^TS_1D_1}
\end{equation*}
for both $\xi = 2,F$. (See Lemma~\ref{lem:mainlemma} below.) Next, we will provide a bound on $\sigma_k^{-1}\left(V_{k}^T S_1D_1S_2\right)$. In order to get a strong accuracy guarantee for the overall algorithm, the deterministic column selection algorithm must satisfy
\begin{equation*}
\sigma_{k}\left(V_k^T S_1 D_1 S_2\right) \geq \frac{\sigma_{k}\left(V_k^T S_1 D_1\right)}{p(k,c)} > 0   ,
\end{equation*}
where $p(k,c)$ is a polynomial in both $k$ and $c$.
Thus, for our main theorem, we will employ Algorithm 4~\cite{GE96} with $f=\sqrt{2}$, which guarantees the above bound with $p(k,c) = \sqrt{2k\left(c-k\right)+1}$. (See Lemma~\ref{lemma:sigmabound} below.) Finally, we will show, using relatively straightforward matrix perturbation techniques, that $\XNorm{\Sigma_{\rho - k} V_{\rho - k}^TS_1D_1}$ is not too much more, in a multiplicative sense, than $\XNorm{A-A_k}$, where we note that the factors differ for $\xi=2,F$. (See Lemmas~\ref{lemma:twonorm} and~\ref{lemma:Fnorm} below.) By combining these results, the main theorem will follow.


\subsection{The rank of $V_{k}^T S_1D_1$}

The following lemma provides a bound on the singular values of the matrix $V_k^TS_1D_1$ computed by the \emph{randomized phase} of Algorithm~\ref{alg:main}, from which it will follow that the matrix
$V_k^TS_1D_1$ is full rank. To prove the lemma, we employ Theorem~\ref{thm:theorem7correct} of the Appendix (this theorem is a variant of a result of Rudelson and Vershynin in~\cite{RV07}). Note that probabilities of the form of eqn.~(\ref{eqn:sampling_probs_firstterm}) actually suffice to establish Lemma~\ref{lemma:rank}.
\begin{lemma}\label{lemma:rank}
Let $S_1$ and $D_1$ be constructed using Algorithm~\ref{alg:main}. Then, with probability at least~$0.9$,
$$
\sigma_k\left(V_{k}^T S_1D_1\right) \geq 1/2   .
$$
In particular, $V_{k}^T S_1D_1$ has full rank.
\end{lemma}
\begin{Proof}
In order to bound $\sigma_k\left(V_{k}^T S_1D_1\right)$, we will bound $\TNorm{V_{k}^T S_1D_1D_1S_1^TV_{k} - I_k}$. Towards that end, we will use Theorem \ref{thm:theorem7correct} with $\beta = 1/2$ and $\epsilon=1/20$, which results in the value for $c$ in eqn.~(\ref{eqn:valueofc}). Note that the sampling probabilities in eqn. (\ref{eq:probexpression}) satisfy
$$p_i \geq \frac{\TNormS{\left(V_k\right)_{(i)}}}{2k}.$$
Now Theorem \ref{thm:theorem7correct} and our construction of $S_1$ and $D_1$ guarantee that for $c$ as in eqn.~(\ref{eqn:valueofc})
$$\Expect{\TNorm{V_k^TV_k-V_k^TS_1D_1D_1S_1^TV_k}}\leq 1/20.$$
We note here that the condition $c_0^2 \FNormS{V_k}\geq 4\beta \epsilon^2$ in Theorem \ref{thm:theorem7correct} is trivially satisfied assuming that $c_0$ is at least one (given our choices for $\beta$, $\epsilon$, and $\FNormS{V_k} = k \geq 1$). Using $V_k^TV_k = I_k$ and Markov's inequality we get that with probability at least~$0.9$,
\begin{eqnarray*}
\nonumber
\TNorm{V_{k}^T S_1D_1D_1S_1^TV_{k} - I_k}
   &\leq& 10\left(1/20\right)=1/2.
\end{eqnarray*}
Standard matrix perturbation theory results~\cite{GVL89} now imply
that for all $i = 1,\ldots,k$,
\begin{equation*}
\label{eqn:spectral1}
\abs{\sigma_{i}^2\left(V_{k}^T S_1D_1\right)-1}
   \leq 1/2.
\end{equation*}
\end{Proof}

\subsection{Bounding the spectral and Frobenius norms of $A-P_CA$}

\begin{lemma}\label{lem:mainlemma}
Let $S_1$, $D_1$, and $S_2$ be constructed as described in Algorithm~\ref{alg:main} and recall that $C = AS_1S_2$. If $V_k^T S_1 D_1$ has full rank, then for $\xi = 2,F$,
\begin{equation*}
\XNorm{A - P_CA} \leq \XNorm{A - A_k} + \sigma_k^{-1}\left(V_{k}^T S_1D_1S_2\right)\XNorm{\Sigma_{\rho - k} V_{\rho - k}^TS_1D_1}.
\end{equation*}
\end{lemma}

\begin{Proof}
We seek to bound the spectral and Frobenius norms of $A - P_CA$,
where $C = AS_1S_2$ is constructed by Algorithm~\ref{alg:main}. To
do so, first notice that scaling the columns of a matrix
(equivalently, post-multiplying the matrix by a diagonal matrix)
by any non-zero scale factors does not change the subspace spanned
by the columns of the matrix. Thus,
\begin{eqnarray}
\nonumber
A - P_CA &=& A - \left(AS_1S_2\right)\left(AS_1S_2\right)^+A         \\
\nonumber
         &=& A - \left(AS_1D_1S_2\right)\left(AS_1D_1S_2\right)^+A    \\
\label{eqn:mainlemma:eq10}
         &=& A - \left(AS\right)\left(AS\right)^+A    ,
\end{eqnarray}
where, in the last line,  we have introduced the convenient
notation $S=S_1D_1S_2 \in R^{n \times k}$ that we will
use throughout the remainder of this proof. In the sequel we seek
to bound the residual
\begin{eqnarray} \label{eqn:tmp}
\XNorm{A - P_CA} = \XNorm{A -
\left(AS\right)\left(AS\right)^+A}.
\end{eqnarray}
First, note that
$$ (AS)^+A = \arg \min_{X \in R^{k \times n}} \XNorm{A - ASX }.$$
This implies that in eqn.~(\ref{eqn:tmp}) we can replace
$(AS)^+A$ with any other $k \times n$ matrix and the
equality with an inequality. In particular we replace
$(AS)^+A$ with $(A_kS)^+A_k$, where $A_k$ is
the best rank-$k$ approximation to $A$:
\begin{eqnarray}
\nonumber \XNorm{A - P_CA}
   &=& \XNorm{ A - AS (AS)^+A } \\
\nonumber   &\leq&  \XNorm{ A - AS (A_kS)^+A_k
}.
\end{eqnarray}
Let $A_{\rho-k} = U_{\rho-k} \Sigma_{\rho-k} V_{\rho-k}^T$. Then,
$A = A_k + A_{\rho-k}$ and, using the triangle inequality,
\begin{eqnarray}
\nonumber \XNorm{A - P_CA} \label{eqn:mainlemma:eq30}
   &=& \XNorm{ A_k + A_{\rho-k} - (A_k + A_{\rho-k})S (A_kS)^+A_k } \\
   \label{eqn:mainlemma:eq40}
   &\leq& \underbrace{\XNorm{ A_k - A_k S (A_kS)^+A_k }}_{\gamma_1} + \underbrace{\XNorm{A_{\rho-k}}}_{\gamma_2} +
   \underbrace{\XNorm{A_{\rho-k}S(A_kS)^+A_k}}_{\gamma_3}.
\end{eqnarray}
We now bound $\gamma_1$, $\gamma_2$, and $\gamma_3$.
First, for $\gamma_1$, note that:
\begin{eqnarray}
\nonumber \gamma_1
   &=& \XNorm{ A_k - A_k S (A_kS)^+A_k } \\
\nonumber
   &=& \XNorm{ U_k \Sigma_k V_k^T - U_k \Sigma_k (V_k^T S) (U_k \Sigma_k V_k^T S)^+U_k \Sigma_k V_k^T}\\ \label{eq3}
   &=& \XNorm{ U_k \Sigma_k V_k^T - U_k \Sigma_k (V_k^T S) (V_k^T S)^+ (U_k \Sigma_k)^+ U_k \Sigma_k V_k^T}\\ \label{eq4}
   &=& \XNorm{\Sigma_k  -  \Sigma_k (V_k^T S) (V_k^T S)^+ (U_k \Sigma_k)^+ U_k \Sigma_k }\\ \label{eq5}
   &=& \XNorm{\Sigma_k  -  \Sigma_k } = 0.\\ \nonumber
\end{eqnarray}
In eqn.~(\ref{eq3}), we replaced $(U_k \Sigma_k V_k^T S)^+$ by $(V_k^T S)^+ (U_k \Sigma_k)^+$. This follows since the statement of our lemma assumes that the matrix $V_k^TS_1D_1$ has full rank. Also, the construction of $S_2$ guarantees that the columns of $V_k^TS_1D_1$ that are selected in the second stage of Algorithm~\ref{alg:main} are linearly independent, and thus the $k \times k$ matrix $V_k^TS= V_k^TS_1D_1S_2$ has full rank and is invertible. In eqn.~(\ref{eq4}), $U_k$ and $V_k^T$ can be dropped without increasing a unitarily invariant norm, while eqn.~(\ref{eq5}) follows since
$V_k^TS$ is a full-rank $k\times k$ matrix. Next, note that $ \gamma_2 = \XNorm{A_{\rho-k}} = \XNorm{A - A_k}$. Finally, to conclude the proof, we bound $\gamma_3$ as follows:
\begin{eqnarray}
\nonumber \gamma_3 &=& \XNorm{A_{\rho-k}S (A_kS)^+A_k} \\
\nonumber &=& \XNorm{U_{\rho-k}\Sigma_{\rho-k} V_{\rho-k}^T  S (U_k \Sigma_k V_k^TS)^+U_k \Sigma_k V_k^T}\\
\label{eq10} &=& \XNorm{\Sigma_{\rho-k} V_{\rho-k}^T  S (V_k^TS)^+}\\
\label{eq12} &\leq& \XNorm{ \Sigma_{\rho - k} V_{\rho - k}^T S}\TNorm{\left(V_{k}^T S\right)^{-1}}\\
\label{eq14}
&=&\sigma_k^{-1}\left(V_{k}^TS\right)\XNorm{\Sigma_{\rho
- k} V_{\rho - k}^TS_1D_1}.
\end{eqnarray}
Eqn.~(\ref{eq10}) follows by the orthogonality of $U_{\rho-k}$ and $V_k$ and the fact that $V_k^TS$ is a $k \times k$ invertible matrix (see above). Eqn.~(\ref{eq12}) follows from the fact that for any two matrices $X$ and $Y$ and $\xi = 2,F$, $\XNorm{XY} \leq \XNorm{X} \TNorm{Y}$. Finally, eqn.~(\ref{eq14}) follows since $S = S_1DS_2$ and $S_2$ is an orthogonal matrix.
\end{Proof}

\subsection{Upper bounds for $\sigma_k^{-1}\left(V_{k}^T S_1D_1S_2\right)$ and $\XNorm{\Sigma_{\rho - k} V_{\rho - k}^TS_1D_1}$, $\xi = 2,F$}

\begin{lemma}\label{lemma:sigmabound}
Let $S_1$, $D_1$, and $S_2$ be constructed using Algorithm~\ref{alg:main}. Then, with probability at least~$0.9$,
$$ \sigma_k^{-1}\left(V_{k}^T S_1D_1S_2\right) \leq 2\sqrt{2k\left(c-k\right)+1}.$$
\end{lemma}

\begin{Proof}
From Lemma \ref{lemma:rank} we know that $\sigma_{i}\left(V_{k}^T S_1D_1\right) \geq 1/2$ holds for all $i=1,\ldots,k$ with probability at least~$0.9$. The deterministic construction of
$S_2$ (see Algorithm 4 of~\cite{GE96} with the parameter $f$ set to $\sqrt{2}$) guarantees that
\begin{eqnarray*}
\sigma_k(V_k^TS_1D_1S_2) \geq \frac{\sigma_k(V_k^TS_1D_1)}{\sqrt{2k\left(c-k\right)+1}} \geq \frac{1}{2\sqrt{2k\left(c-k\right)+1}}.
\end{eqnarray*}
\end{Proof}

\begin{lemma}\label{lemma:twonorm}\textbf{($\xi = 2$)}
If $S_1$ and $D_1$ are constructed as described in~Algorithm \ref{alg:main}, then, with probability at least~$0.9$,
\begin{equation*}
\TNorm{\Sigma_{\rho - k} V_{\rho - k}^TS_1D_1}  \leq \TNorm{A-A_k} + \frac{4}{c^{1/4}}\FNorm{A-A_k}.
\end{equation*}
\end{lemma}
\begin{Proof}
Let $\Gamma = \Sigma_{\rho-k}V_{\rho - k}^TV_{\rho - k}\Sigma_{\rho-k} = \Sigma_{\rho-k}^2$. We manipulate $\TNormS{\Sigma_{\rho - k} V_{\rho - k}^TS_1D_1}$ as follows:
\begin{eqnarray*}
\TNormS{\Sigma_{\rho-k} V_{\rho -k}^T S_1D_1}
   &=& \TNorm{\Sigma_{\rho-k}V_{\rho - k}^T S_1D_1 D_1S_1^TV_{\rho - k}\Sigma_{\rho-k}}  \\
   &=& ||\Sigma_{\rho-k}V_{\rho -k}^T S_1D_1 D_1S_1^T V_{\rho - k}\Sigma_{\rho-k}
     - \Gamma
     + \Gamma||_2  \\
   &\leq& ||\Sigma_{\rho-k}V_{\rho - k}^T S_1D_1D_1S_1^T V_{\rho - k}\Sigma_{\rho-k}
     - \Gamma||_2 + \TNorm{\Sigma_{\rho-k}^2} \\
  &\leq& ||\Sigma_{\rho-k}V_{\rho - k}^T S_1D_1D_1S_1^T V_{\rho - k}\Sigma_{\rho-k}
     - \Gamma||_F + \TNorm{\Sigma_{\rho-k}^2}.
\end{eqnarray*}
Given our construction of $S_1$ and $D_1$ and applying eqn. (9) of Theorem 1 of \cite{dkm_matrix1} with $\beta = 1/2$ and $\delta=0.1$, we get
that with probability at least~$0.9$,
\begin{eqnarray*}
\FNorm{\Sigma_{\rho-k}V_{\rho - k}^T S_1D_1D_1S_1^T V_{\rho - k}\Sigma_{\rho-k}-
\Sigma_{\rho-k}V_{\rho - k}^TV_{\rho - k}\Sigma_{\rho-k}}
\leq \frac{12}{\sqrt{c}}\FNormS{\Sigma_{\rho-k}V_{\rho-k}^T}.
\end{eqnarray*}
Thus, by combining the above results and using $\TNorm{\Sigma_{\rho-k}^2} = \TNormS{A-A_k}$ and
$\FNormS{\Sigma_{\rho-k}V_{\rho-k}^T} = \FNormS{A-A_k}$ we get
\begin{equation*}
\TNormS{\Sigma_{\rho-k} V_{\rho -k}^T S_1D_1}
   \leq
   \frac{12}{\sqrt{c}}\FNormS{A-A_k}
   + \TNormS{A-A_k}.
\end{equation*}
To conclude the proof of the lemma we take the square roots of both sides of the above inequality.
\end{Proof}

\begin{lemma}
\label{lemma:Fnorm}
\textbf{($\xi = F$)}
If $S_1$ and $D_1$ are constructed as described in~Algorithm \ref{alg:main}, then, with probability at least~$0.9$, $$ \FNorm{\Sigma_{\rho - k} V_{\rho - k}^TS_1D_1} \leq 4\FNorm{A - A_k}. $$
\end{lemma}
\begin{Proof}
It is straightforward to prove that with our construction of $S_1$ and $D_1$, the expectation of $\FNormS{\Sigma_{\rho - k} V_{\rho - k}^TS_1D_1}$ is equal to $\FNormS{\Sigma_{\rho - k} V_{\rho - k}^T}$. In addition, note that the latter quantity is exactly equal to $\FNormS{A-A_k}$. Applying Markov's inequality, we get that, with probability at least~$0.9$,
$$\FNormS{\Sigma_{\rho - k} V_{\rho - k}^TS_1D_1} \leq 10\FNormS{A - A_k}.$$
Taking square roots of both sides of the above inequality concludes the proof of the lemma.
\end{Proof}

\subsection{Completing the proof of Theorem~\ref{thm:main}}

To prove the Frobenius norm bound of Theorem~\ref{thm:main} we combine Lemma~\ref{lem:mainlemma} (with $\xi=F$) with Lemmas~\ref{lemma:sigmabound} and~\ref{lemma:Fnorm}. Thus, we get
\begin{eqnarray*}
\FNorm{A - P_CA} &\leq& \FNorm{A - A_k} + \left(2\sqrt{2k\left(c-k\right)+1}\right)\left(4\FNorm{A - A_k}\right)\\
&=& \left(1+ 8\sqrt{2k\left(c-k\right)+1}\right)\FNorm{A - A_k}.
\end{eqnarray*}
Using $c = \Theta\left(k \log k\right)$ immediately derives the Frobenius norm bound of Theorem~\ref{thm:main}. Notice that Lemma~\ref{lemma:sigmabound} fails with probability at most $0.1$ and that Lemma~\ref{lemma:Fnorm} fails with probability at most $0.1$; thus, applying the standard union bound, it follows that the Frobenius norm bound of Theorem~\ref{thm:main} holds with probability at least $0.8$. To prove the spectral norm bound of Theorem~\ref{thm:main} we combine Lemma~\ref{lem:mainlemma} (with $\xi=2$) with Lemmas~\ref{lemma:sigmabound} and~\ref{lemma:twonorm}. Thus, we get
\begin{eqnarray*}
\TNorm{A - P_CA} &\leq& \TNorm{A - A_k} + \left(2\sqrt{2k\left(c-k\right)+1}\right)\left(\TNorm{A-A_k} + \frac{4}{c^{1/4}}\FNorm{A-A_k}\right)\\
&=& \left(1+ 2\sqrt{2k\left(c-k\right)+1}\right)\TNorm{A - A_k}+\frac{8\sqrt{2k\left(c-k\right)+1}}{c^{1/4}}\FNorm{A-A_k}.
\end{eqnarray*}
Using $c = \Theta\left(k \log k\right)$ immediately derives the spectral norm bound of Theorem~\ref{thm:main}.  Notice that Lemma~\ref{lemma:sigmabound} fails with probability at most $0.1$ and that Lemma~\ref{lemma:twonorm} fails with probability at most $0.1$; thus, applying the standard union bound, it follows that the spectral norm bound of Theorem~\ref{thm:main} holds with probability at least $0.8$.
\section*{Acknowledgements} We are grateful to Daniel Spielman and Ilse Ipsen for numerous useful discussions on the results of this paper. We would also like to thank an anonymous reviewer of an earlier version of this manuscript who provided a counterexample to Lemma 4.4 of~\cite{BMD09_CSSP_SODA}, and thus helped us identify the error in the proof of that lemma.


\section*{Appendix}

Let $A \in \mathbb{R}^{m \times n}$ be any matrix. Consider the following algorithm (which is essentially the algorithm in page 876 of~\cite{DMM08_CURtheory_JRNL}) that constructs a matrix $C\in \mathbb{R}^{m \times c}$ consisting of $c$ rescaled columns of $A$. We state Theorem 4 of~\cite{DMMS07_FastL2_TR} that provides a bound for the approximation error $\TNorm{AA^T-CC^T}$.

\begin{algorithm}[ht]
\begin{framed}

\SetLine

\AlgData{
$A \in \mathbb{R}^{m \times n}$,
$p_i \geq 0, i\in[n]$ s.t. $\sum_{i \in [n]}p_i=1$,
positive integer $c \leq n$.}

\AlgResult{
$C \in \mathbb{R}^{m \times c}$
}

Initialize $S \in \mathbb{R}^{m \times c}$ to be an all-zero matrix.

\For{$t=1,\ldots,c$}{
   Pick $i_t \in [n]$, where $\textbf{Prob}\left(i_t = i\right) = p_i$\;
   $S_{i_t t} = 1/\sqrt{cp_{i_t}}$\;
}

Return $C = AS$\;

\end{framed}
\caption{
The \textsc{Exactly($c$)} algorithm.
}
\label{alg:SDconstruct_exact}
\end{algorithm}

\begin{theorem}\label{thm:theorem7correct}
Let $A \in \mathbb{R}^{m \times n}$ with $\TNorm{A} \leq 1$. Construct $C$ using the \textsc{Exactly($c$)} algorithm and let the sampling probabilities $p_i$ satisfy
\begin{equation}\label{eqn:defPj}
p_i \geq \beta \frac{\TNormS{A^{(i)}}}{\FNormS{A}}
\end{equation}
for all $i \in [n]$ for some constant $\beta \in (0,1]$. Let $\epsilon \in (0,1)$ be an accuracy parameter, assume $c_0^2 \FNormS{A}\geq 4\beta \epsilon^2$, and let
\begin{equation*}\label{eqn:Cbound}
c = 2 \left(\frac{c_0^2 \FNormS{A}}{\beta \epsilon^2}\right)\log \left(\frac{c_0^2 \FNormS{A}}{\beta \epsilon^2}\right).
\end{equation*}
(Here $c_0$ is the unknown constant of Theorem 3.1, p. 8 of~\cite{RV07}.) Then,
$$\Expect{\TNorm{AA^T-CC^T}}\leq \epsilon.$$
\end{theorem}

\end{document}